# Toward a Comprehensive Model of Snow Crystal Growth:
# 6. Ice Attachment Kinetics near -5 C

Kenneth G. Libbrecht

Department of Physics
California Institute of Technology
Pasadena, California 91125
kgl@caltech.edu

**Abstract. I examine a variety of snow crystal growth measurements taken at a temperature of -5 C, as a function of supersaturation, background gas pressure, and crystal morphology. Both plate-like and columnar prismatic forms are observed under different conditions at this temperature, along with a diverse collection of complex dendritic structures. The observations can all be reasonably understood using a single comprehensive physical model for the basal and prism attachment kinetics, together with particle diffusion of water vapor through the surrounding medium and other well-understood physical processes. A critical model feature is structure-dependent attachment kinetics (SDAK), for which the molecular attachment kinetics on a faceted surface depend strongly on the nearby mesoscopic structure of the crystal.**

## 1. INTRODUCTION

Ever since Nakaya, Hanajima, and Mugurama first described how snow crystal growth morphologies vary with temperature and supersaturation [1954Nak, 1942Han, 1944Han, 1958Nak], researchers have struggled to explain these fairly basic observations. Numerous experimental studies have confirmed and extended the earlier results [1958Hal, 1961Kob, 1990Yok, 2009Bai, 2012Bai], and the observations are often summarized in the *snow crystal morphology diagram* [2017Lib, 2019Lib]. Devising a comprehensive physical model to explain the full diversity of growth behaviors, however, has proven to be a remarkably challenging task.

The two dominant processes that determine snow crystal growth dynamics are particle diffusion (typically the diffusion of water vapor through a background gas of air) and the molecular attachment kinetics at the ice surface. It is useful to describe the latter in terms of the Hertz-Knudsen relation [1882Her, 1915Knu, 1996Sai, 1990Yok, 2005Lib, 2017Lib, 2019Lib]

$$v_n = \alpha v_{kin} \sigma_{surf} \qquad (1)$$

where $v_n$ is the crystal growth velocity normal to a growing surface, $\alpha$ is a dimensionless *attachment coefficient*, $\sigma_{surf}$ is the water vapor supersaturation at the surface, and $v_{kin}$ is a *kinetic velocity* that incorporates the statistical mechanics of ideal gases. A detailed discussion of this equation and its foundations can be found in [2019Lib].

In addition to particle diffusion and attachment kinetics, there are numerous additional physical processes that affect ice growth dynamics to a lesser degree, including surface-energy effects, notably the Gibbs-Thomson effect, thermal diffusion and latent heating, and possibly chemical effects at the ice surface. As described in [2019Lib], these processes are generally less important than the dominant processes of particle diffusion and attachment kinetics, and this will be true in the discussion below as well.

While the physics of particle diffusion is well understood and calculable, this is not the case for the attachment kinetics. How $\alpha$ depends on temperature, $\sigma_{surf}$, and other parameters depends on the detailed many-body physics that describes how water vapor molecules become incorporated into the ice crystal lattice. Our relatively poor understanding of the ice attachment kinetics at present thus reflects our generally poor understanding of the ice surface structure and the complex molecular dynamics that takes place on growing ice/vapor interfaces.

On molecularly "rough" ice surfaces, it appears that impinging water vapor molecules are essentially immediately incorporated into the solid structure, meaning $\alpha \approx 1$ is a good approximation on such surfaces. There is no compelling experimental evidence that goes against this statement (to my knowledge), so I will assume $\alpha \approx 1$ on rough ice surfaces in what follows.

The attachment kinetics is profoundly different on the basal and prism faceted surfaces, however, where often $\alpha \ll 1$. More than any other physical effect, this anisotropy in the attachment kinetics is responsible for the formation of macroscopic faceted surfaces during snow crystal growth. In particular, the anisotropy in the attachment kinetics is far more important than the surface-energy anisotropy, which appears to be essentially negligible in snow crystal growth dynamics [2019Lib].

To a large degree, therefore, the development of a comprehensive model of snow crystal growth depends critically on developing a physically realistic molecular model of $\alpha_{basal}$ and $\alpha_{prism}$, the attachment kinetics on the basal and prism facets as a function of $\sigma_{surf}$, the growth temperature $T$, and other factors. I have attempted to create such a model in [2019Lib1], and in this paper I find that the model is quite well supported by numerous ice-growth measurements at -5 C. Moreover, this close examination of a broad range of data provides many useful insights into the essential physics underlying the ice attachment kinetics.

I am hopeful, therefore, that the model presented in [2019Lib1] may represent a substantial step forward toward developing a comprehensive model of snow crystal growth dynamics. At the very least, the measurements presented here should help provide a framework for discussing further investigations in pursuit of the long-awaited explanation of the snow crystal morphology diagram.

## 2. A Physical Model of the Ice Attachment Kinetics

In this section I describe the physical origins and overarching features of my proposed model of the ice/vapor attachment kinetics near -5 C, defining $\alpha_{basal}$ and $\alpha_{prism}$ over a wide range of growth conditions. This will then lead into a detailed comparison with ice-growth experiments in the next section.

Because our theoretical understanding of crystal growth dynamics is relatively primitive, especially for the case of ice, one cannot simply write down a fully formed model of the attachment kinetics from first principles. Instead, experimental measurements and theoretical inference must go hand in hand to develop a self-consistent picture of the relevant underlying physical processes. Nevertheless, for the sake of pedagogy, I believe it is beneficial to first describe the model characteristics in detail, largely without justification. Once the overall picture is presented, we can then proceed to justifying the model with experimental observations.

Figure 1 illustrates the attachment coefficients $\alpha_{basal}$ and $\alpha_{prism}$ as a function of $\sigma_{surf}$ over a broad range of growth conditions when the growth temperature is near -5 C. The full temperature-dependent model of the attachment kinetics is described in [2019Lib1], and Figure 1 shows a detailed look at this model for a fixed temperature of -5 C. Note that single-valued functions $\alpha_{basal}(\sigma_{surf})$ and $\alpha_{prism}(\sigma_{surf})$ are not sufficient to encompass all aspects of ice/vapor growth near -5 C, so several different "branches" of this (semi-empirical) theory are illustrated in Figures 1 and 2, requiring a fair bit of supporting discussion.

## Basal Attachment Kinetics

Consider first the blue curve labeled "large basal surface" in Figure 1, which represents the growth of a broad faceted basal surface. This curve applies to the limiting case of a basal surface of infinite lateral extent, where edge-dependent structural effects are irrelevant. Here the attachment kinetics is well described by the nucleation and growth of terrace steps, for which there is an established theoretical description in most crystal-growth textbooks [1996Sai, 1999Pim, 2002Mut]. Assuming a polynucleation model in classical nucleation theory from the vapor phase, I write the attachment coefficient as

$$\alpha_{basal}(\sigma_{surf}) = A_{basal} e^{-\sigma_{0,basal}/\sigma_{surf}} \quad (2)$$

where $\sigma_{0,basal}$ is a parameter that derives from the terrace step energy on the faceted surface and $A_{basal}$ depends on the admolecule surface diffusion and other parameters. In classical nucleation theory, $A_{basal}$ may exhibit a weak dependence on $\sigma_{surf}$, but this dependence is negligibly small in the current discussion, given the substantial uncertainties in experimental data, so I assume that $A_{basal}$ is essentially a constant parameter. Measurements indicate (see below) that $\sigma_{0,basal} \approx 0.007 = 0.7\%$ and $A_{basal} \approx 1$ so these values were used to define the large-basal-surface curve in Figure 1.

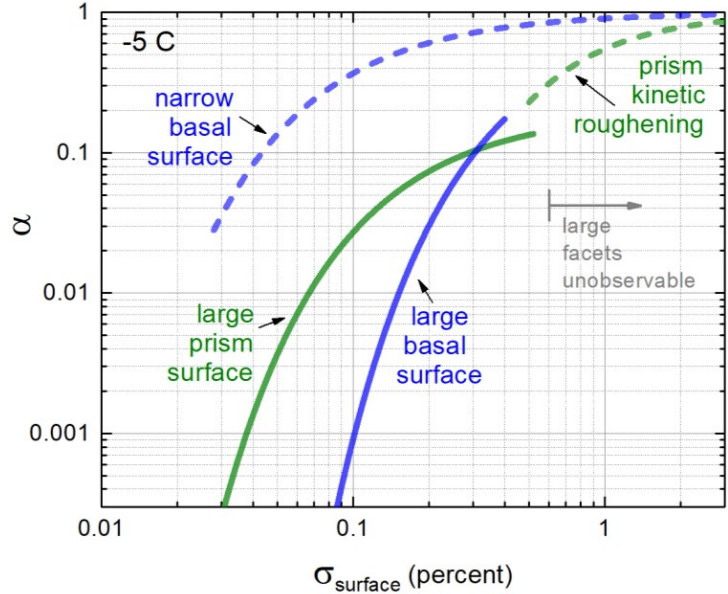

**Figure 1:** The proposed model of the ice/vapor attachment coefficient on basal and prism surfaces when the growth temperature is near -5 C, adapted from [2019Lib1]. The physical significances of the different curves are discussed in detail in the text.

I ended the large-basal-surface curve at $\sigma_{surf} \approx 0.4\%$ in Figure 1 because there is, at present, no known experimental method for accurately measuring $\alpha_{basal}$ on large faceted surfaces at supersaturations much above this level. When $\sigma_{surf}$ is appreciably higher than 0.4% at -5 C, the crystal growth rate is so rapid that thermal and particle diffusion become dominant factors, limiting the growth so strongly that $\alpha_{basal}$ becomes unmeasurable. (Or, more accurately, one cannot distinguish $\alpha_{basal}$ from unity.) It appears reasonable, at this point, to assume that the large-basal-surface curve in Figure 1 extends to $\sigma_{surf} > 0.4\%$ without additional modification, but there appears to be no easy way to determine whether this is actually the case. (Terminating the curve also reduces clutter in the figure, which is useful.)

A key feature in the model in [2019Lib1] is the phenomenon of *Structure Dependent Attachment Kinetics* (SDAK), in which the basal attachment kinetics at -5 C depend strongly on the overall mesoscopic structure of the crystal. In particular, $\alpha_{basal}$ increases markedly when the width of the top basal terrace becomes sufficiently narrow. In Figure 1, the "narrow basal surface" curve refers to the narrow edge of a hollow-column crystal or the sharp tip of a c-axis needle crystal, where the radius of curvature of the surface may be of order $R_{basal} \approx 1\ \mu m$. The narrow-basal-surface curve in Figure 1 is defined by Equation (2) with $A_{basal} = 1$ and $\sigma_{0,basal} = 0.1\%$, but this is meant to represent a rough approximation of reality, and even the functional form of this curve is not well known. The essential feature in Figure 1 is that $\alpha_{basal}$ is much larger on narrow basal surfaces than on broad basal surfaces, and I describe a possible physical mechanism that produces this behavior in [2019Lib1].

Note that measurements together with crystal-growth theory indicate that a surface curvature of $R \approx 1\ \mu m$ is somewhat characteristic for the tips of dendritic snow crystals growing in air [2002Lib, 2017Lib, 2019Lib]. Thus the narrow-basal-surface curve in Figure 1 should apply reasonably well to many snow crystal structures growing in air at -5 C. That being said, "narrow" is a somewhat subjective term in this context. The actual value of $\alpha_{basal}$ for a particular crystal may lie anywhere between the two basal curves in Figure 1, depending on the width of the basal surface. For a quantitative model, one should therefore write $\sigma_{0,basal} = \sigma_{0,basal}(R_{basal})$, where the $R_{basal}$ dependence is not well defined at present [2015Lib2], requiring additional experimental and/or theoretical input. While this deficiency renders the model somewhat incomplete and ill-determined, this is the best we can do at present.

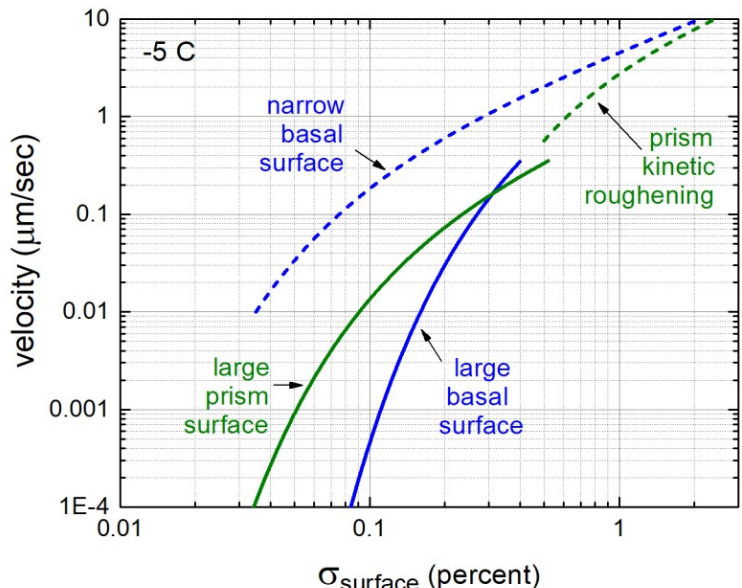

**Figure 2: The model in Figure 1 converted to growth velocities using Equation (1).**

The physical origin of the SDAK effect, specifically that $\sigma_{0,basal}$ depends on $R_{basal}$ as illustrated in Figure 1, was described in detailed in [2019Lib1], at -5 C involving the surface diffusion of admolecules from the edges of basal facets. The SDAK model is a novel idea and is certainly not well established at present. Indeed, the main purpose of the present investigation is to test the model as thoroughly as possible by comparing it with ice growth measurements. I chose -5 C for this purpose because the SDAK model makes a number of unusual predictions at this temperature. In particular, the model predicts the formation of plate-like crystals at -5 C, a behavior that is counter to the well-established snow crystal morphology diagram. The clear observations of plate-like forms (below) suggests that the SDAK phenomenon plays an important role in snow crystal growth, being essential for explaining the snow crystal morphology diagram [2019Lib1]. The narrow-basal-surface curve in Figure 1 encapsulates the SDAK phenomenon at -5 C, and I will examine its effects in detail when comparing the model with experiments below.

## Prism Attachment Kinetics

Turning our attention to the prism facet, the large-prism-surface curve in Figure 1 again describes basic nucleation-limited growth with a functional form equivalent to that in Equation (2). Here again, measurements have determined $\sigma_{0,prism} \approx 0.2\%$ and $A_{prism} \approx 0.2$ near -5 C [2013Lib], and these values have been used in Figure 1. As with basal faceted surfaces, the rapid growth when $\sigma_{surf} > 0.5\%$ makes measurement of the attachment kinetics of large faceted prism surfaces impractical at high supersaturations.

It is instructive to consider why $A_{prism} < 1$ while $A_{basal} \approx 1$. According to the SDAK model presented in [2019Lib1], the "SDAK dip" occurs at quite low temperatures on prism surfaces, implying that surface premelting is quite substantially developed on the prism facet at -5 C, more so than on the basal facet. Assuming a fairly thick quasi-liquid layer (QLL) on the prism surface, I propose that $A_{prism} < 1$ means that the lateral growth of prism terraces at the ice/QLL interface is somehow frustrated by the dynamics of molecular movement at that interface. Normal surface diffusion (on a non-premelted solid/vapor interface) is hindered by the QLL, while the premelted layer is still too thin at -5 C to allow liquid-like bulk diffusion. While certainly speculative, it seems physically plausible that some kind of QLL effect could yield $A_{prism} < 1$ on prism surfaces in the presence of significant surface premelting. That the basic molecular character of premelting looks different on the basal and prism facets has been proposed before [1982Kur, 2019Lib1], but the underlying physics of this phenomenon is not well understood.

All the QLL-related aspects of this model of the attachment kinetics are quite uncertain, fundamentally because the dynamics of crystal growth at a crystal/QLL/vapor interface is not generally well understood. Indeed, the growth dynamics at solid/liquid interfaces are already less tractable compared to solid/vapor interfaces, and premelting adds additional unknown complexities. Nevertheless, the measured $A_{prism} < 1$ does seem to suggest that the ice growth dynamics at an ice/QLL interface is somehow impeded on prism surfaces, more so than on basal surfaces, so I adopt that as a working hypothesis for now.

According to [2019Lib1], the SDAK effect should not be present on prism surfaces at -5 C, so there is no narrow-prism-surface curve analogous to the narrow-basal-surface curve in Figure 1. Nevertheless, the data presented below suggest that $\alpha_{prism} \rightarrow 1$ at sufficiently high $\sigma_{surf}$, which is inconsistent with $A_{prism} \approx 0.2$ measured at lower $\sigma_{surf}$.

Speculating once again, I propose that the impediment to the lateral growth of prism terraces that causes $A_{prism} < 1$ becomes unimportant when new terraces are nucleated at a sufficiently high rate, thereby yielding many island terraces in close proximity. This is essentially the phenomenon of "kinetic roughening" [1996Sai, 1999Pim, 2002Mut], where terrace nucleation is so high that the surface becomes effectively rough, in this case yielding $\alpha_{prism} \rightarrow 1$ at very high $\sigma_{surf}$. The prism-kinetic-roughening curve in Figure 1 is little more than a rough guess at present, based mainly on the limited data presented below. Once again, this region of parameter space is quite difficult to explore using ice-growth experiments.

## A Work in Progress

Clearly the model in Figure 1 is more complicated than one would like, with an uncomfortably large component of speculation as to the relevant physical processes taking place at the basal and prism surfaces under different conditions. It appears, however, that snow crystal growth is a remarkably complex phenomenon, with peculiar behaviors observed as a function of temperature, supersaturation, and other parameters like surface mesostructure. The model in Figure 1 and [2019Lib1] was developed to explain the full panoply of experimental observations, including a diverse range of growth behaviors that probably cannot be fully comprehended with any simpler model.

To add to the difficulty of our task, ice-growth experiments have often produced contradictory results, especially when comparing experiments done at different times by different researchers. It is necessary, therefore, to choose which experimental data to accept and which to reject. I describe this process a bit further below, and additional critical examination of some older experiments can be found in [2004Lib]. The study of snow crystal growth dynamics is very much a work in progress, so part of our job is develop an overarching picture of the fundamental physical processes involved using necessarily incomplete and sometimes erroneous inputs.

As I hope to show below, the model in Figure 1 is in reasonable agreement with a wide range of experimental results, making it superior to previous attempts to model the ice/vapor attachment kinetics. Moreover, its physical underpinnings are quite plausible (in my opinion). Hopefully, this represents real progress. Combining these results with the broader picture in [2019Lib1], I believe this represents a significant step toward developing a comprehensive physical model of snow crystal growth. However, my goal here is not to have the final word on this subject, but to continue developing a sensible framework for describing the quantitative analysis of snow crystal growth dynamics, which will hopefully stimulate additional work and development over time.

### Background Gas Pressure

Note that the model in Figure 1, which is essentially a subset of the variable-temperature model in [2019Lib1], does not include any dependence of the attachment kinetics on background gas pressure, at least for air or other similarly inert gases. As described in [2019Lib], the question of how background gases affect snow crystal growth has been debated in the literature for many years, with no definitive resolution. Particle and heat diffusion are clearly mediated by background gas, and many chemically active gases produce profound changes in snow-crystal growth behaviors, even in quite small concentrations. For air and other inert gases, however, it is not entirely certain as to whether any surface-specific processes, such as attachment kinetics, terrace nucleation, or admolecule surface diffusion, are significantly affected by background-gas interactions.

The evidence generally suggests that interactions with air and similar background gases have little effect on the ice/vapor attachment kinetics. For example, ice growth observations in air, nitrogen, helium, argon, hydrogen, carbon dioxide, and methane gases at a pressure of one bar have yielded similar crystal morphologies as a function of temperature [1959Heu, 2011Lib, 2008Lib1], suggesting that these gases are not sufficiently chemically active to modify the attachment kinetics. The observations do not preclude some kind of gas interactions that significantly change the attachment kinetics in clean air, but none has been definitively demonstrated to date. More work in this area is needed, but in this paper I assume that background gas pressure (for air and other inert gases) has no direct effect on the ice/vapor attachment kinetics. I also present some additional evidence to this effect below.

## 3. Comparisons with Ice Growth Data

Having defined the quantitative model in Figure 1, we are now ready to compare it with relevant experimental data. Our main focus will be on precise ice growth measurements that allow us to extract useful information on the attachment kinetics, which requires that we also manage a host of additional physical effects that arise in the measurement process, for example from particle diffusion, latent heating, and surface-energy effects.

Our overarching strategy will be to measure crystal growth velocities and use Equation (1) to extract $\alpha_{basal}$ and $\alpha_{prism}$ as a function of $\sigma_{surf}$ and other parameters. Our principal difficulty is that $\sigma_{surf}$ cannot be measured directly, but must be determined from other experimental means. This presents measurement challenges that have plagued many ice-growth experiments [2004Lib], and we have only recently come to fully appreciate the potential systematic errors that can result from improper compensation for these effects.

## Low Pressure Data 1

We begin with the set of experimental data shown in Figure 3, taken by Libbrecht and Rickerby using the apparatus described in [2013Lib]. I believe that this experiment provided one of the most precise measurements of the ice/vapor attachment kinetics, superseding previous efforts, as it covered a wide range in $T$ and $\sigma_{surf}$ while carefully managing a variety of systematic experimental effects. Some characteristics of this experiment include:

1) These growth measurements were made with a background gas pressure at or below 0.03 bar, greatly reducing particle diffusion effects. Under these conditions, $\sigma_{surf}$ was quite close to $\sigma_\infty$, the latter being the supersaturation far from a growing crystal. Small residual differences between $\sigma_{surf}$ and $\sigma_\infty$ were modeled and subtracted.

2) A plane-parallel growth chamber was used, with the upper plate (the ice reservoir) being just 1 mm above the lower plate (the test-crystal substrate). Moreover, the typical spacing between the test crystal and any neighboring crystals was >1 mm. As described in [2019Lib], this geometry minimizes large-scale diffusion effects that often reduce $\sigma_\infty$ from the theoretical value one calculates using the temperature difference between the ice reservoir and the test crystal substrate. This systematic error can be extremely large if not recognized and carefully managed [2015Lib].

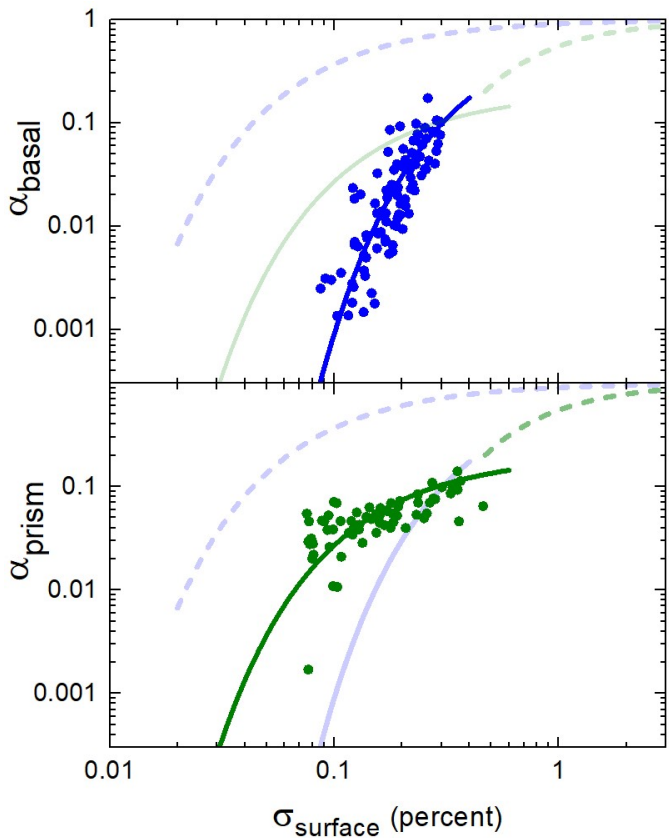

Figure 3: Measurements of $\alpha_{basal}$ (top, blue data points) and $\alpha_{prism}$ (bottom, green data points) as a function of $\sigma_{surf}$, at a growth temperature of $T \approx -5\,C$. The data are from the experiment described in [2013Lib], taken at background air pressures at or below 0.03 bar. The lines are reproduced from Figure 1.

3) By measuring small crystals on a sapphire substrate, heating effects were largely negligible for the basal growth measurements and quite small for the prism growth measurements.

4) The crystals were large enough (typically about 20 $\mu m$) that the Gibbs-Thomson effect was negligible, but small enough that residual heating and diffusion effects were manageable.

5) Substrate interactions were avoided by measuring the growth of faceted surfaces aligned parallel to the substrate, which thus did not intersect the substrate. This avoids issues relating to substrate-initiated terrace nucleation [2019Lib].

6) White-light interferometry and laser interferometry were used make precise measurements of growth velocities.

7) A large number of crystals were measured to examine crystal-to-crystal variation and various systematic effects in the experiment.

We found that careful attention to these and other experimental details was essential for obtaining accurate and physically meaningful results. Our choice of $A_{basal}$, $A_{prism}$, $\sigma_{0,basal}$ and $\sigma_{0,prism}$ in Figure 1 was informed largely from the experimental data shown in Figure 3.

The finite experimental ranges in $\sigma_{surf}$ in this experiment were mainly determined by two factors: 1) below $\sigma_{surf} \approx 0.1\%$, it was difficult to determine $\sigma_{surf}$ with high accuracy, as this quantity was derived from the small temperature difference between the substrate and reservoir plates; and 2) at high $\sigma_{surf}$, the growth rates were so large that heating and particle diffusion corrections became unmanageably large.

## Substrate Growth in Air 1

Ice growth in air at normal pressure is strongly limited by particle diffusion, a fact that makes it difficult to obtain useful information about the attachment kinetics from growth measurements in air [2019Lib]. A relevant parameter is

$$\alpha_{diff} = \frac{X_0}{R} = \frac{c_{sat}}{c_{ice}} \frac{D}{v_{kin}} \frac{1}{R} \qquad (3)$$

where $R$ is the overall crystal size and $X_0 \approx 0.14 \, \mu m$ in air at one bar. If $\alpha_{diff} \ll \alpha$, then the growth will be so strongly diffusion-limited that little information about $\alpha$ can be obtained from measured growth velocities [2019Lib]. As a result, absolute measurements of $\alpha$ in air are limited to regions where $\alpha$ is quite small, which usually means operating at small $\sigma_{surf}$.

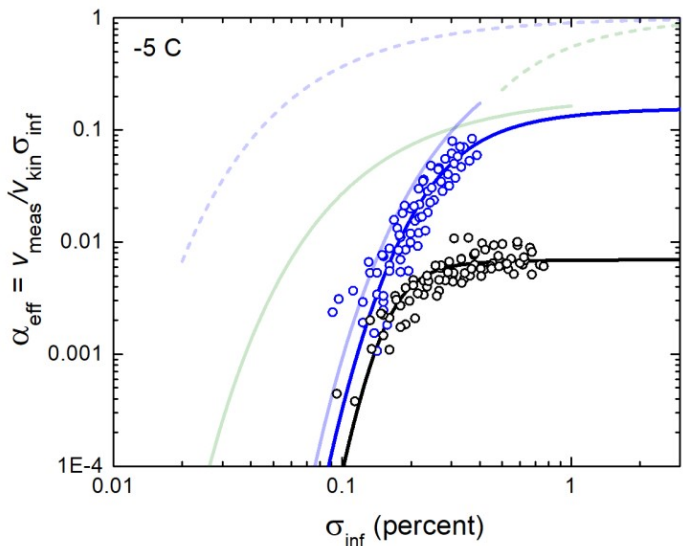

**Figure 4:** The (upper) blue data points show measurements of $\alpha_{eff} = v_n/v_{kin}\sigma_\infty$ for basal growth at 0.03 bar. These are the same basal data shown in Figure 3, excepted here uncorrected for residual particle-diffusion effects. The heavy blue line shows a model of the uncorrected data [2012Lib]. The (lower) black points show similar data taken at 1 bar, exhibiting much larger particle diffusion effects. At low growth rates, both sets of data agree with a terrace nucleation model with $\sigma_{0,basal} \approx 0.007 = 0.7\%$, supporting the SDAK model at -5 C [2013Lib2].

Figure 4 illustrates one informative example for basal growth in air near -5 C. When the air pressure is low (blue points in Figure 4), then the measured quantity $\alpha_{eff} = v_n/v_{kin}\sigma_\infty$ will be approximately equal to $\alpha_{basal}$. Thus the blue data points in Figure 4 were used, after correcting for small residual diffusion effects, to obtain the $\alpha_{basal}(\sigma_{surf})$ measurements shown in Figure 3. For the blue data in Figure 4, $\alpha_{diff} \approx 0.2$ and the blue line shows a model of $\alpha_{eff}$ that incorporates diffusion effects. At high $\sigma_{surf}$, $\alpha_{diff} < \alpha_{basal}$ and it becomes impossible to accurate correct the data to obtain $\alpha_{basal}(\sigma_{surf})$.

The black data points in Figure 4 were obtained from measurements in air at one bar, where $\alpha_{diff} \approx 0.007$. In this case the model line shows that the measurements were strongly limited by diffusion effects except at the lowest values of $\sigma_{surf}$. The fact that the two sets of data converge at low $\sigma_{surf}$ suggests that, at least in this low-$\sigma_{surf}$ regime, $\alpha_{basal}(\sigma_{surf})$ is roughly independent of background air pressure.

The main take-away from the data in Figure 4 is that $\sigma_{0,basal}$ in air cannot be substantially lower than $\sigma_{0,basal}$ at low pressure, for the case of large basal surfaces. This important bit of information was also incorporated into the model in Figure 1, as the data suggest that large basal surfaces exhibit nucleation-limited growth that is independent of air pressure. This contrasts the much higher $\alpha_{basal}$, and thus lower $\sigma_{0,basal}$, exhibited on narrow basal surfaces, which is the central tenet of the SDAK phenomenon. Differentiating between SDAK effects and possible air-pressure-dependent effects is a subtle point, but it is important for developing a self-consistent model of the attachment kinetics that agrees with observations.

## Low Pressure Data 2

Because precision measurements like those in Figures 3 and 4 are so important for understanding the ice attachment kinetics, I recently developed a new ice-growth experiment to corroborate these results [2020Lib]. The new apparatus again uses a parallel-plate geometry, but now a single pulse from a close-proximity expansion nucleator [2019Lib] randomly deposits a collection of minute ice crystals onto the substrate for simple imaging of their subsequent growth. While the resulting $v_n$ measurements not as precise as the interferometric measurements in [2013Lib], the new experiment provides a higher throughput with smaller test crystals, in a cleaner environment with reduced systematic effects from particle diffusion and other factors. Figure 5 shows some low-pressure data from this new experiment for growth at -5 C.

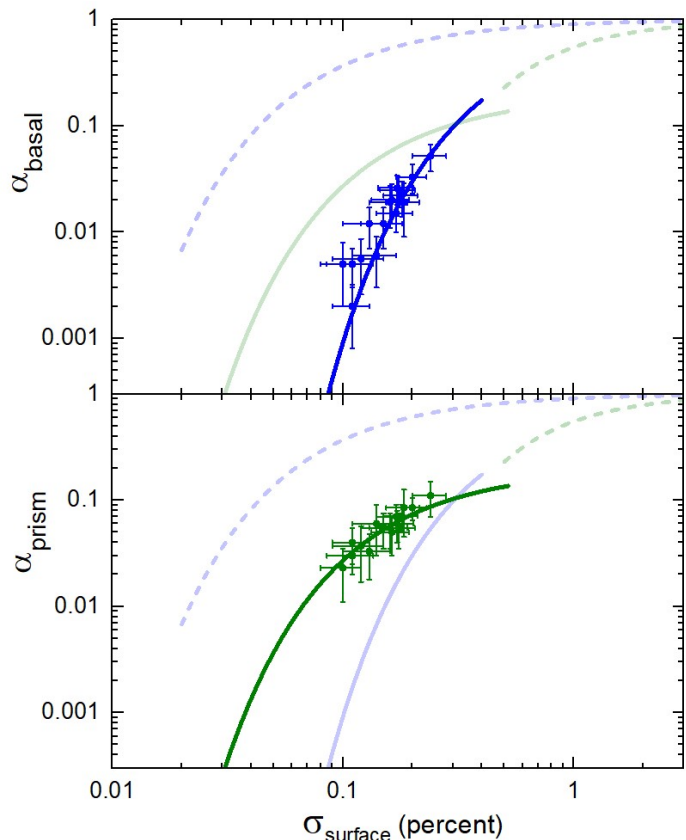

Figure 5: Additional measurements of $\alpha_{basal}$ (top, blue) and $\alpha_{prism}$ (bottom, green) as a function of $\sigma_{surf}$ at a growth temperature of $T \approx -5\ C$ and a background air pressure of 0.1 bar. These data were obtained using a new substrate-growth experiment described in [2020Lib], and the lines are reproduced from Figure 1. The new data are in good agreement with the older data shown in Figure 3.

Special attention was given to substrate interactions in these experiments, using a thoroughly cleaned sapphire substrate with a fresh hydrophobic surface coating applied before each run. The coating yielded contact angles of >90 degrees for water droplets at room temperature, and the ice/substrate contact angles at -5 C are likely comparable or greater. With this surface treatment, it appears that the rate of substrate-induced terrace nucleation is quite low [2019Lib], although perhaps not yet entirely negligible.

The main result to date from these measurements is that the new data confirm the previous measurements to a large degree, as can be seen by comparing Figures 3 and 5. While supporting the model in Figure 1 for broad-facet growth, the new data directly confirm the growth of plate-like simple prisms at -5 C when $\sigma_{surf}$ is low, as this morphological behavior is unambiguously observed in the imaging data. This direct observation of plate-like prismatic crystals at -5 C provides an important confirmation of our attachment-kinetics model. Producing columnar crystals at -5 C then requires the SDAK effect to increase $\alpha_{basal}$ compared to its value on broad basal facets.

Thick plates were found to be the norm when $\sigma_{surf} \approx 0.15\%$, with prism aspect ratios of roughly $\rho_{aspect} = H/R \approx 0.5$, where $H$ is the half-thickness and $R$ is the effective radius of the hexagonal prism. As described in [2020Lib], $\rho_{aspect}$ is not always a good indicator of $\alpha_{basal}/\alpha_{prism}$, however, as $\sigma_{surf}$ generally becomes smaller as a crystal grows. Thus $\rho_{aspect}$ depends on the entire growth history of a particular crystal, and detailed modeling is required to measure $\alpha_{basal}$ and $\alpha_{prism}$ with high accuracy [2020Lib].

Supporting data from this new experiment also suggest that substrate interactions still produce some terrace nucleation events on basal surfaces, as these surfaces intersect the substrate at 90 degrees. It may be difficult to reduce the residual rate of substrate-induced terrace nucleation to zero, given that surface hydrophobic coatings cannot be made perfectly uniform, and this would prevent $\alpha_{basal}$ from dropping as precipitously as expected at low $\sigma_{surf}$. Better coatings, especially superhydrophobic coatings, could largely eliminate this problem, but this awaits additional development work.

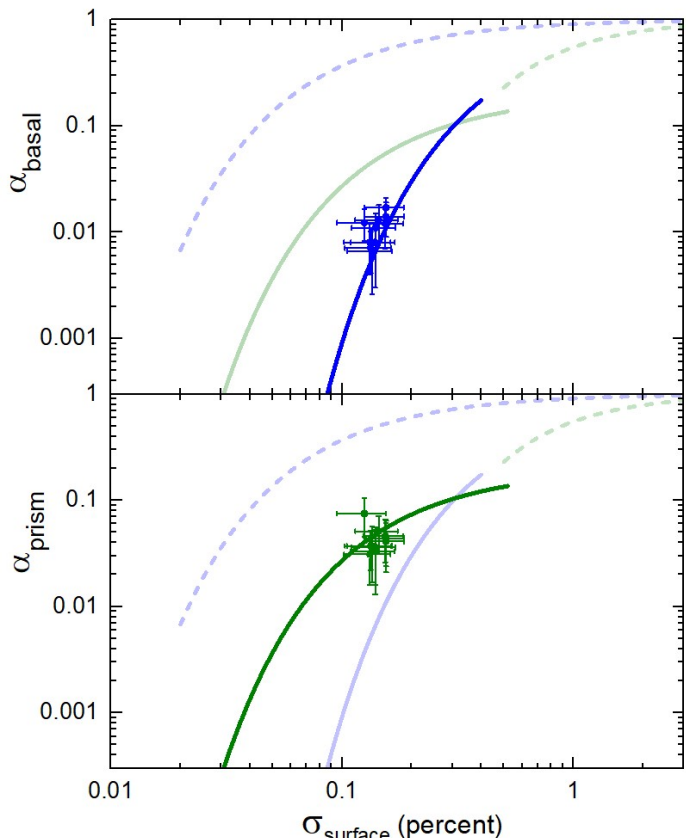

Figure 6: Measurements of $\alpha_{basal}$ (top, blue) and $\alpha_{prism}$ (bottom, green) as a function of $\sigma_{surf}$ at a growth temperature of $T \approx -5\,C$ and a background air pressure of 1 bar. These data were obtained using the same apparatus used for the data in Figure 5 [2020Lib], and the lines are reproduced from Figure 1. Once again, the data suggest that the attachment kinetics are largely unaffected as the air pressure increases from 0.1 to 1 bar.

When the supersaturation was increased to $\sigma_{surf} \approx 0.3\%$, nucleated prisms grew with essentially isometric morphologies having $\rho_{aspect} \approx 1$, as expected from the model in Figure 1. I have not yet been able to further test the model by observing prismatic columnar forms at $\sigma_{surf} > 0.4\%$, as would be expected. The primary difficulty at high $\sigma_{surf}$ is that thermal effects become quite large in this region of parameter space, as latent heat generated at the crystal surface is not conducted to the substrate at a sufficient rate to prevent significant heating at the crystal extremities. This problem becomes especially severe at high growth rates, as the crystal size increases rapidly, exacerbating these unwanted thermal effects. For this reason, as indicated in Figure 1, it becomes practically impossible to observe simple faceted growth at high $\sigma_{surf}$.

## Substrate Growth in Air 2

Particle diffusion strongly limits growth when $\alpha_{diff} \ll \alpha$, so little useful information can be gleaned about $\alpha$ from ice growth measurements when this inequality holds. However, diffusion effects can be effectively modeled for $\alpha_{diff}$ values as low as $\alpha_{diff} \approx \alpha$. This means that useful measurements of $\alpha(\sigma_{surf})$ in air can be obtained only if sufficiently small crystals are used. For crystals with $R > 3\,\mu m$, which is something of a practical limit with our optical imaging, this means restricting observations to approximately $\alpha < 0.05$.

Figure 6 shows some additional growth data taken at a pressure of 1 bar, which can be compared with the similar data in Figure 5 taken at 0.1 bar. In both cases, the direct imaging data are in reasonable agreement with the interferometric measurements in Figure 3 [2013Lib]. Barring not-entirely-negligible systematic effects (see [2020Lib]) from residual diffusion modeling, substrate interactions, Gibbs-Thomson corrections (which are significant when working with especially small crystals at low $\sigma_{surf}$), and thermal corrections (important at high $\sigma_{surf}$), the data from both these experiments are in reasonable agreement, supporting the model presented in Figure 1.

At this juncture I would like to again call attention to the fact that the model in Figure 1 predicts plate-like growth for prismatic crystals at low $\sigma_{surf}$. This prediction contradicts a well-established rule from the morphology diagram, suggesting that only columnar crystals are produced at -5 C. More precisely, numerous earlier studies of snow crystal morphologies have indicated aspect ratios that varied monotonically from $\rho_{aspect} \gg 1$ at high $\sigma_\infty$ to $\rho_{aspect} \approx 1$ at low $\sigma_\infty$ in air [2019Lib]. However, there have been some reports of plate-like crystals growing at -5 C [1980Kel, 2012Kni], and I have found that the model in Figure 1 does a reasonable job of making sense of these observations as well. I will elaborate further on this point as I consider the various experiments below.

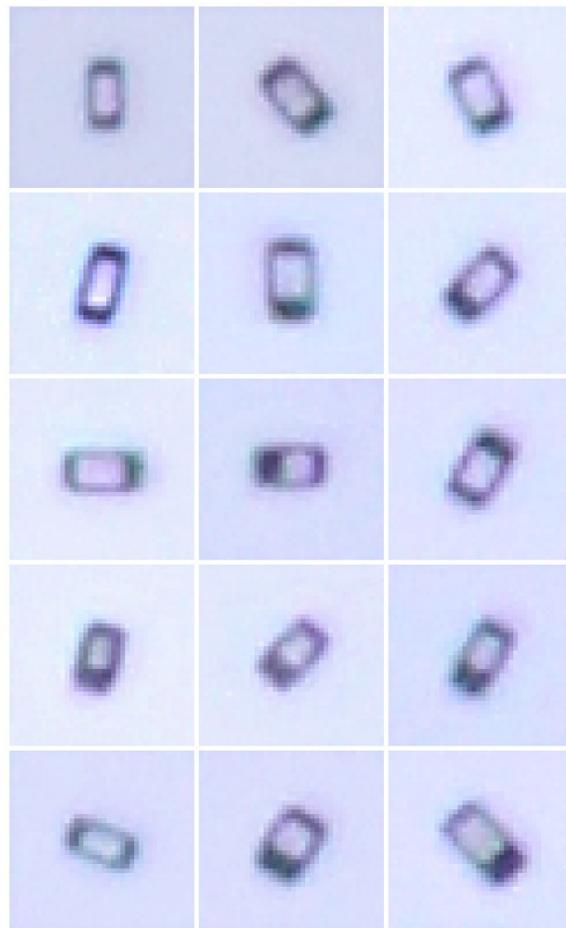

**Figure 7:** A representative sample of simple prisms growing on a substrate in air at -5 C, with a far-away supersaturation of $\sigma_\infty \approx 0.2\%$. and a (modeled) near-surface supersaturation of $\sigma_{surf} \approx 0.12\%$. Measured crystal sizes from this sample are shown in Figure 8.

The model in Figure 1 certainly predicts columnar growth with $\rho_{aspect} \gg 1$, but this behavior arises largely from the SDAK phenomenon [2019Lib1], which yields the narrow-basal-surface line in Figure 1. In contrast, simple prismatic growth follows the large-basal-surface line in Figure 1, indicating plate-like growth with $\rho_{aspect} < 1$ for sufficiently low $\sigma_{surf}$. In many prior morphological studies, it appears that this region of parameter space was not carefully explored, so the transition from columnar to plate-like growth was not observed.

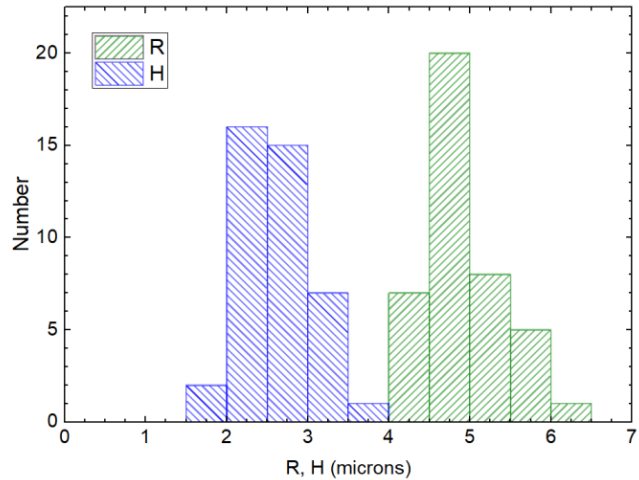

**Figure 8:** A histogram of crystal effective radius (R) and half-thickness (H) from a sample that included the crystals shown in Figure 7. These data included all well-formed simple prisms from the sample, indicating that thick-plates were the norm while columnar crystals were entirely absent. The measured areal density on the substrate was ~15 crystals per square millimeter (total), of which ~20% were sufficiently well-formed to be included in this graph. The crystals not included were mostly just oriented at odd angles with respect to the substrate.

Our model is nicely supported by the new direct-imaging experiment [2020Lib], as thick-plate prisms are the norm for growth in air at low growth rates when the crystal sizes are small. Figure 7 illustrates some representative examples from a single run, and Figure 8 shows a histogram of crystal sizes from this run. Note that Figure 8 includes all well-formed crystals from this particular sample, meaning clean prismatic samples where the c-axis of the crystal was parallel to the substrate. (Radial sizes of crystals with the c-axis perpendicular to the substrate were similar. Oddly oriented crystals were consistent with thick-plate growth, but clean measurements of both R and H were only possible when the c-axis was parallel to the substrate.)

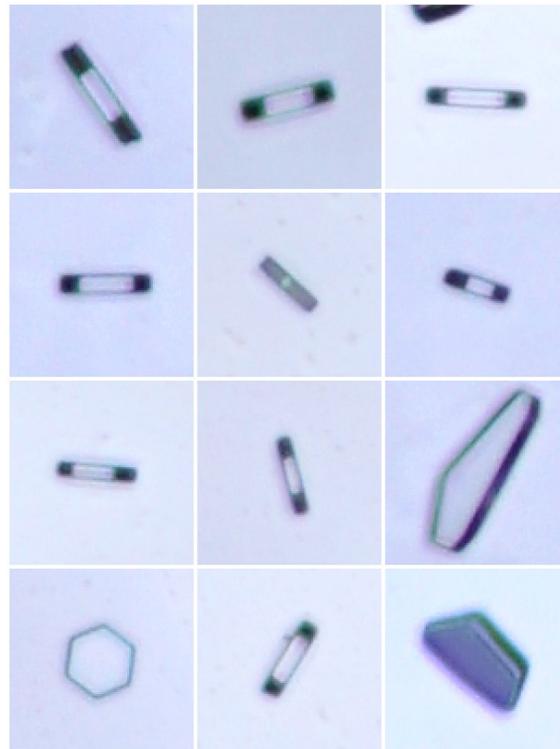

**Figure 9:** Selected examples of thin-plate prismatic growth at -5 C, observed when the supersaturation was especially low. These selected crystals are *not* representative of the entire sample, however, suggesting that residual substrate interactions may have nucleated higher-than-natural basal growth on many crystals, yielding blockier prisms in addition to some thin plates. This suggests that thin-plate crystals would be the normal growth form at -5 C and low supersaturations, if substrate interactions could be avoided.

Independent of air pressure, thick-plate crystals were readily observed at low $\sigma_{surf}$ in many trials, while columnar crystals were absent. Barring any large and unknown systematic effects, these highly reproducible observations clearly show that plate-like prisms are the normal growth form for -5 C at low $\sigma_{surf}$. Below I discuss how this observation fits together with the many long-standing observations of columnar prisms at -5 C.

As the supersaturation is reduced, Figure 1 predicts that $\rho_{aspect}$ should become smaller, yielding quite thin plate-like crystals at very low $\sigma_{surf}$. While it is difficult to nucleate crystals at extremely low $\sigma_{surf}$ (because small crystals are unstable to sublimation via the Gibbs-Thomson effect [2019Lib]), these conditions can be obtained by nucleating a crowded field of crystals on a substrate and simply letting them grow larger, as diffusion effects then readily lower $\sigma_{surf}$.

Doing this experiment yielded mixed results, producing a diversity of prismatic crystals with aspect ratios throughout the range $0.1 < \rho_{aspect} < 1$. Columnar crystals (with $\rho_{aspect} > 1$) were absent, and Figure 9 shows some examples of thin-plate crystals that formed. These crystals serve as a proof-of-principle only, demonstrating that thin-plate crystals can form at -5 C when $\sigma_{surf}$ is low. But thin plates like these were not the norm in our sample, as many blockier, nearly isometric crystals were also present. Put another way, the sample distribution (in terms of $\rho_{aspect}$) was not sharply peaked, in contrast to the quite narrow distributions found with smaller crystals at higher $\sigma_{surf}$, as shown in Figures 7 and 8.

I believe that contact with the substrate may be playing a significant role in these observations, and my working hypothesis at present is that there is typically a low level basal-terrace nucleation brought about by substrate interactions. When the basal growth rate is in the 5-10 nm/sec range (as for the crystals in Figure 7 and 8), the substrate nucleation rate is largely negligible. But this substrate effect can be important when the basal growth rate drops below 1 nm/sec (for the crystals in Figure 9). Moreover, because the substrate surface coating is not perfectly uniform, one expects a fairly broad spread in observed $\rho_{aspect}$ values, as is observed. In this picture, the crystals shown in Figure 9 represent cases where the hydrophobic coating was especially effective, so substrate-induced basal nucleation is mostly absent. Additional work, especially with levitated or freely falling crystals at low $\sigma_{surf}$, will be needed to verify that plate-like crystals with especially low $\rho_{aspect}$ values are indeed the norm at the lowest supersaturations, as predicted in the model in Figure 1.

## Free Fall Growth in Air

Figure 10 presents another relevant data set taken at -5 C, showing measurements of crystals that have grown in free-fall through supersaturated air at 1 bar [2009Lib, 2019Lib]. In essence, a vessel of heated water at the bottom of a one-meter-tall growth chamber yielded supersaturated air as convection mixed the evaporated water vapor with the air in the chamber. The resulting supersaturation was somewhat ill-determined and non-uniform, but Figure 10 shows $\sigma_\infty$ values measured using differential hygrometry at the center of the chamber. The crystal measurements themselves were somewhat biased as well, selecting well-formed columnar crystals and rejecting blockier morphologies. Columnar crystals were the most likely form observed, but information pertaining to the entire $\rho_{aspect}$ distribution was not recorded.

In our previous analysis [2009Lib], we assumed that the measured $\sigma_\infty$ values were correct, but I now believe that the actual supersaturation was somewhat lower. The main culprit was the removal of water vapor by the growing crystals themselves, which can be a surprisingly large effect unless the overall density of crystals in the chamber is kept quite low. The actual $\sigma_\infty$ then depends on how quickly the lost water vapor is replenished by convection and diffusion, which is not possible to determine with good accuracy.

To proceed, therefore, I assume here that the prism growth rates are given by the large-prism-surface curve in Figure 1. This assumption should apply because the prism facets were quite large in extent, while the basal surfaces were smaller (especially on hollow columns) and therefore subject to the SDAK phenomenon. With this assumption, the measured prism growth velocities in Figure 10 can be used to estimate $\sigma_{surf}$ directly from Figure 2, circumventing entirely the need to know $\sigma_\infty$. Making this assumption is tantamount to assuming that the model in Figure 1 correctly describes the prism growth, so the data in Figure 10 can then no longer be thought of as a decisive test of the model. Instead, these data only allow us to examine the SDAK effect on the basal surfaces, and the results serve as something of an overall consistency check, examining how the free-fall measurements can reasonably be interpreted in the context of the proposed model. Put another way, this is an imperfect data set, but we will try to see what we can learn from it nevertheless.

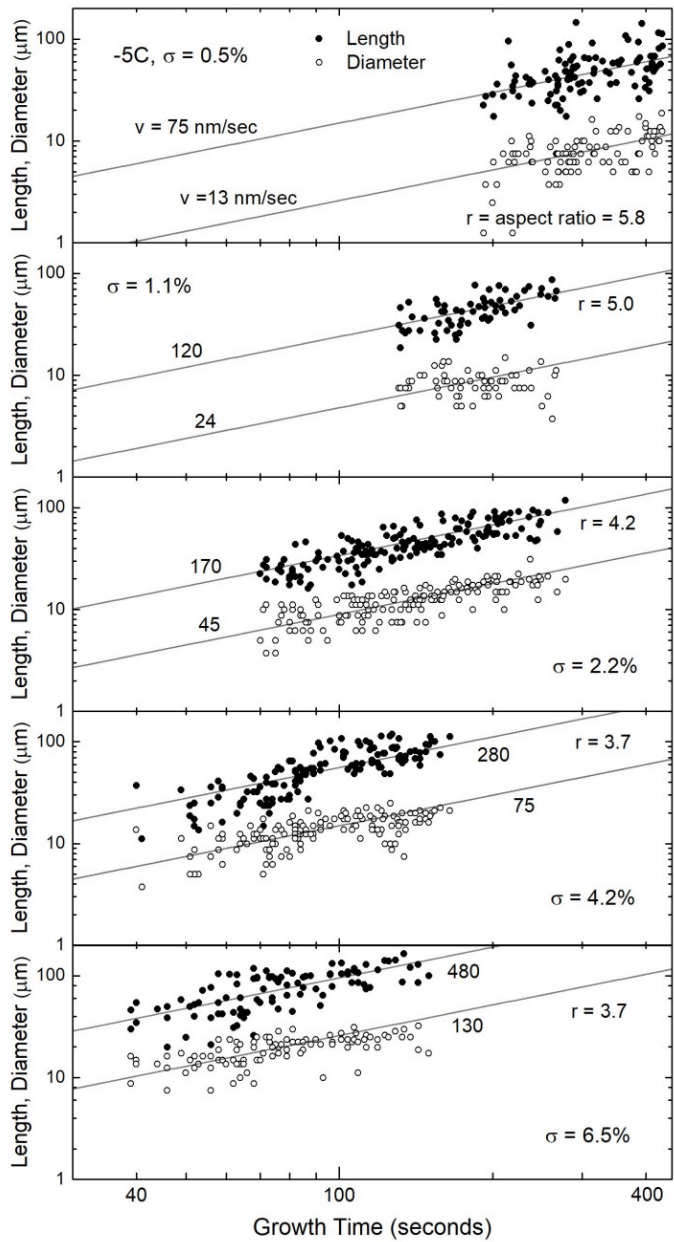

**Figure 10: (Right)** Free-fall growth data at -5 C from [2009Lib], showing crystal sizes as a function of fall times for several different supersaturations (here $\sigma = \sigma_\infty$, which is generally not equal to $\sigma_{surf}$). Each pair of points (length and diameter) represent one observed crystal. Lines show constant-velocity trajectories, and the aspect ratio is the ratio of these velocities. The accompanying numbers show constant-velocity lines in nm/sec. Note that the aspect ratio tends toward unity with increasing supersaturation.

Proceeding with these caveats in mind, we obtain the results shown in Figure 11, providing estimates of both $\sigma_{surf}$ and $\alpha_{basal}$ corresponding to the five panels shown in Figure 10. This is something of a weak result overall, as it basically assumes that the basal SDAK phenomenon is responsible for the columnar growth behavior. But it is useful for examining what additional experiments of this nature would be useful.

Note that the diffusion corrections are so large in these data that it is impossible to determine $\sigma_{surf}$ directly from the measurements. This is best understood from the analytical model for spherical growth [2019Lib], which indicates

$$\delta\sigma = \sigma_\infty - \sigma_{surf}$$
$$\approx \frac{R}{X_0}\frac{v}{v_{kin}} \qquad (4)$$

Using a typical crystal size in Figure 10 as a rough estimate of $R$ in this expression, we quickly find that $\delta\sigma \approx \sigma_\infty$, meaning that the diffusion effects are so large in these data that it is impossible to model $\sigma_{surf}$ directly with a useful accuracy. This fact was not sufficiently appreciated in [2009Lib], and I now believe that the analysis for $\alpha_{basal}$ and $\alpha_{prism}$ in that paper was largely incorrect. Given that realization, the model-dependent analysis shown in Figure 11 is probably the best we can do to interpret these free-fall data.

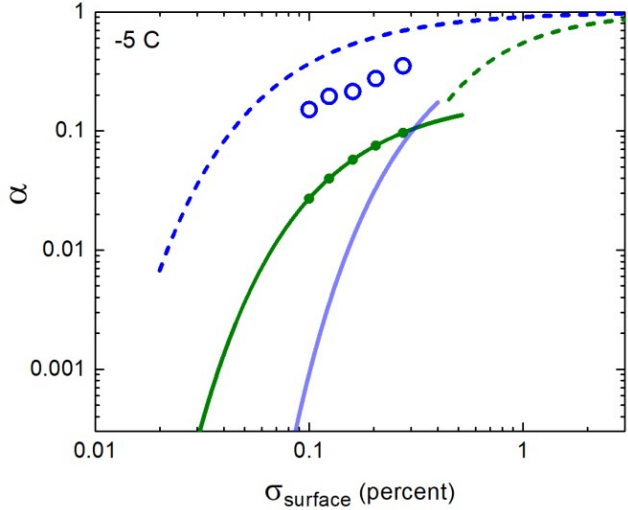

**Figure 11: Comparing the free-fall data in Figure 10 with the model in Figure 1. Here the prism growth velocity was assumed to agree with the model (solid green points), thus determining $\sigma_{surf}$, and this value was assumed to be equal on the basal and prism surfaces. The values of $\alpha_{basal}$ (open blue points) were then calculated from $\sigma_{surf}$ and the measured basal growth velocity.**

That being said, Figure 11 does give us accurate values of $\sigma_{surf}$ if the $\alpha_{prism}$ model is correct, and that can be useful for comparing with other data sets to better understand the behavior of this model. For example, Figures 6 and 11 both display ice growth measurements in air, with similar values of $\sigma_{surf}$ for both sets of data. Yet the growth morphologies are remarkably different, with thick-plate crystals in Figure 6 and slender columnar crystals in Figure 11. Figure 9 presents an even greater contrast, clear demonstrating the existence of thin plates growing at -5 C [1980Kel, 2012Kni]. The different experiments seem to present rather wildly contradictory results (thin plates versus slender columns under ostensibly identical growth conditions), but I believe that both sets of data can be reasonably explained by the model in Figure 1.

For the columnar data in Figures 10 and 11, the crystals were nucleated in an environment with the rather large values of $\sigma_\infty$ shown in Figure 10 (although perhaps a bit lower than these numbers, as mentioned above). When the crystals were first nucleated into this supersaturated air, they were sub-micron in size, and diffusion effects are generally smaller for small crystals. Thus $\sigma_{surf}$ was high for these nascent crystals, perhaps in the 1% range for a brief time, yielding the high growth rates shown in Figure 2. This initially fast growth likely initiated the SDAK effect on the basal surfaces, driving $\alpha_{basal}$ to the narrow-basal-surface curve in Figure 1. In addition to columnar hollowing, a symmetry-breaking instability may have caused one side of the basal surface to sharpen, even if full symmetrical hollowing did not develop.

Because of this early rapid growth, basal sharpening put $\alpha_{basal}$ on the SDAK track, and this fast growth maintained the sharp edge and kept $\alpha_{basal}$ on that track. This is the nature of the edge-sharpening instability (ESI) described in [2019Lib1]. Once initiated, this instability caused the basal surface to keep it sharp-edged morphology, yielding the observed slender columnar crystals. The value of $\sigma_{surf}$ dropped considerably as the crystals grew, owing to diffusion effects, but not enough to halt the SDAK effect. In fact, the SDAK effect may have subsided in some crystals in this free-fall sample, causing them to evolve into blocky forms. But these crystals would have been rejected, as only slender columns were included in the data in Figure 10. As with Figure 9, the displayed data does not represent the complete sample, which included a greater range of morphologies.

In comparison, the crystals in Figures 6, 7, and 9 did not experience an initial spurt of fast growth. The nucleated crystals deposited onto the substrate with few-micron or sub-micron sizes, but the initial supersaturation was low, so $\sigma_{surf}$ was low from the outset. In this environment, the SDAK mechanism could not develop on the basal surfaces, so these surfaces developed into large, slow-growing basal facets, yielding low $\alpha_{basal}$ values.

A critical part of this picture is that the SDAK effect, driving the Edge-Sharpening Instability [2019Lib1], provides a new kind of snow-crystal growth instability, somewhat related to the Mullins-Sekerka branching instability [1963Mul, 1964Mul, 2019Lib], but different in that it involves a meso-structure-dependent change in the attachment kinetics. In principle, one could incorporate the SDAK mechanism into a full numerical model of snow crystal growth, and hopefully the different behaviors observed in these experiments would be reproduced. We are some ways away from accomplishing this goal, but the attachment-kinetics model as it stands does seem to reasonably explain the disparate experimental observations.

If this line of reasoning is indeed correct, it predicts that we should be able to observe columnar growth on a substrate in air at -5 C, provided we just start the experiment with a sufficiently high $\sigma_\infty$ when nucleating crystals. I performed this experiment, and Figure 12 shows some example crystals using $\sigma_\infty \approx 3\%$ (the exact value was somewhat ill-determined, and $\sigma_\infty$ rapidly decreased from diffusion effects as the deposited crystals grew larger). Well-formed columnar crystals do not readily form on a substrate, as the substrate boundary condition disrupts the usual columnar symmetry. Nevertheless, Figure 12 shows that columnar crystals will form in this experiment if the initial supersaturation is relatively high. This was demonstrated in [2013Lib2] also.

Once again, the reader may feel that the model in Figure 1, along with its subsequent explanations and caveats, contains too many epicycles of operational complexity to be considered a viable theory. This may be true, but the challenge at present is to come up with any physically realistic model, simple or complex, that can explain the remarkable diversity of snow crystal forms observed, in this case even at a single temperature of -5 C. The model presented in Figure 1 may not be correct in every detail, but I believe that the ice attachment kinetics will likely require a model of equal or greater complexity to explain the full menagerie of snow crystal growth forms. In particular, it becomes remarkably difficult to explain the full range of observations without some variant of the SDAK phenomenon.

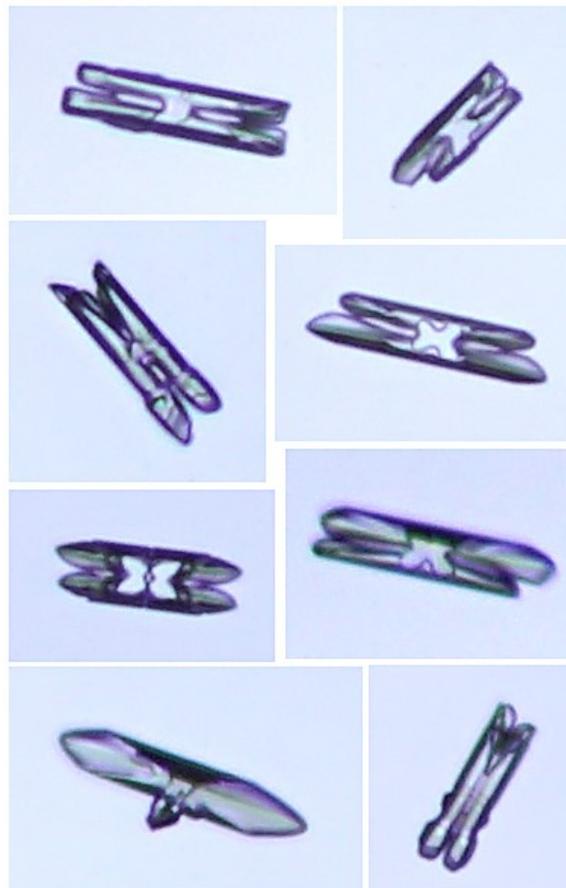

**Figure 12: Examples of columnar crystals growing on a substrate in air at -5 C, obtained using the same apparatus used to obtain the data shown in Figures 5-9 [2020Lib]. These observations (along with the discussion in the text) suggest that a brief interval of fast growth followed by a longer period of slower growth is the usual recipe for growing slender columnar snow crystals in air.**

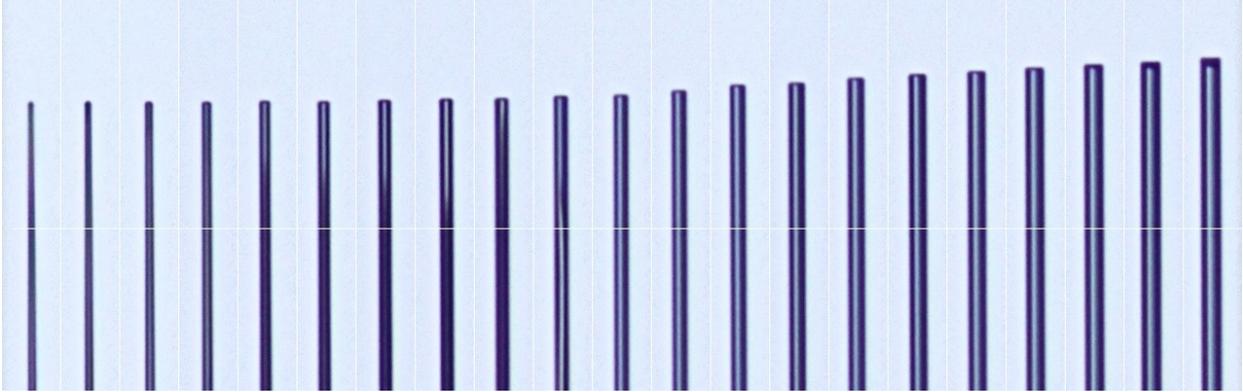

Fortunately, the proposed model makes numerous predictions that can be investigated in future ice-growth experiments. For example, it should be possible to grow thin plates in air at -5 C without the constraints of a substrate, provided the supersaturation is sufficiently low. The growth rates will be exceedingly slow, of order 1-10 nanometers per second, so long growth times will be required to observe these crystals. However, if a suitable experiment can be devised (a nontrivial challenge), then it should be straightforward to grow freely falling or levitated thin plates at -5 C.

### Electric Needles in Air

The growth of "electric" ice needles [2019Lib] provides another interesting experimental system that can be employed to test the kinetics model in Figure 1. Briefly, high electric fields are used to stimulate the growth of c-axis needle crystals in the first of two diffusion chambers, and the needles are then transported to a second diffusion chamber to observe their subsequent growth in normal air with no electric fields, under a variety of environmental conditions [2014Lib1, 2019Lib]. This apparatus allows observations a quite high supersaturation levels, although the growth of the resulting large structures is strongly affected by diffusion effects. Crystal heating has been observed in addition to particle diffusion using this apparatus [2016Lib], but our focus here will be on what we can learn about the attachment kinetics.

**Figure 13:** A composite image showing the growth of a c-axis electric needle in air at **-5 C** with a far-away supersaturation of $\sigma_\infty = 1.8\%$ [2016Lib1]. Midway through this series, the initially tapered needle developed faceted prism surfaces, while the basal growth abruptly increased. Measurements of this behavior are shown in Figure 14.

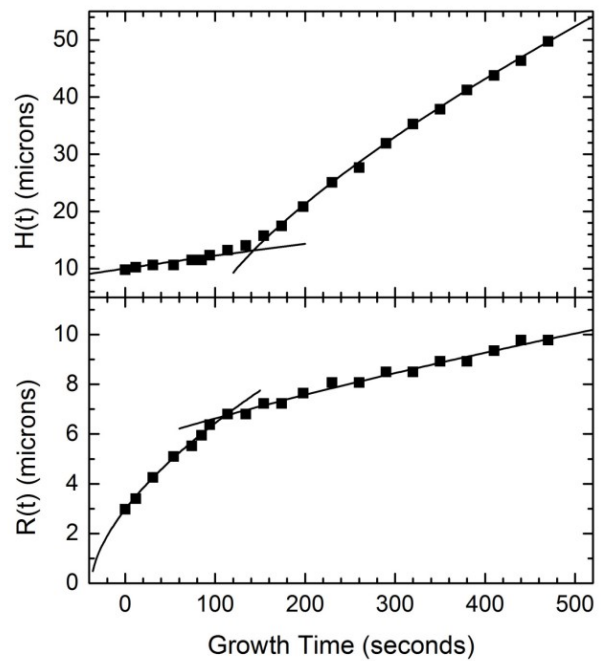

**Figure 14:** Measurements of the needle height $H(t)$ and tip radius $R(t)$ extracted from the images in Figure 13 [2016Lib1]. Lines were drawn to guide the eye, with $(v_{basal}, v_{prism}) \approx (\mathbf{100}, \mathbf{9})$ nm/sec after the transition. A detailed explanation for this unusual growth behavior is given in the text.

Figures 13 and 14 illustrate a particularly interesting time-series of images, in which a needle crystal exhibited a clear growth transition as it developed. At early times, the needle had a tapered structure, so the attachment coefficient on the needle sides was relatively high, owing to the closely spaced series of prism terrace steps on the vicinal surfaces. In this circumstance, the radial growth was relatively high and $\sigma_{surf}$ was relatively low, the growth being strongly limited by particle diffusion. In this regime, the Gibbs-Thomson effect inhibited the basal growth, as the needle tip was quite sharp. Indeed, at much lower values of $\sigma_\infty$, newly created needles exhibit sublimation of the tip, which continues until $R$ is large enough to halt the sublimation.

Two changes happen around $t \approx 130$ seconds in Figure 14 – the needles sides become faceted, and the SDAK effect causes an abrupt increase in $\alpha_{basal}$ at the needle tip. Performing a quantitative analysis of this transition has proven to be quite tricky, as the growth is strongly limited by diffusion, making it nearly impossible to determine $\sigma_{surf}$ accurately. I now believe that the analysis described in [2016Lib1] was flawed, and one cannot fully understand the growth behavior using these observations alone. The fundamental problem is that same as with the freely falling needle crystals in the previous section – when the growth is strongly limited by particle diffusion, it becomes practically impossible to extract information about the attachment kinetics in an absolute sense.

As with the needle crystals above, the only piece of information one can reliably use in these situations is the ratio of the basal and prism growth rates. Therefore, as with the free-fall data, I use the prism growth velocity after $t \approx 130$ in Figure 14 to infer $\sigma_{surf}$ using the model in Figure 1. Then I use the basal growth velocity to extract $\alpha_{basal}$ at that value of $\sigma_{surf}$. This is a crude analysis, and model-dependent as well, but it useful for determining whether

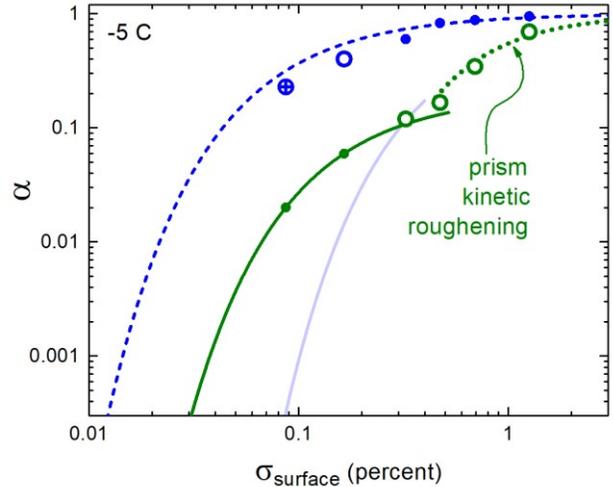

**Figure 15: Observations of the growth of needle, columnar, and dendritic structures growing on the ends of electric needle crystals, in comparison with the model curves from Figure 1. The high-growth data show that $\alpha_{prism}$ deviates from the nucleation-limited behavior seen at lower $\sigma_{surf}$, which I interpret as evidence for kinetic roughening on the prism surfaces at high growth rates.**

the model provides a sensible explanation for these unusual observations.

Performing this analysis gives us the first pair of points in Figure 15, showing $\alpha_{basal}$ and $\alpha_{prism}$ at $\sigma_{surf} = 0.09\%$. Because the overall needle length is quite long, particle diffusion yields quite a low value of $\sigma_{surf}$, accompanied by a prism growth velocity of just 9 nm/second. Even with the low value of $\alpha_{prism}$ shown in Figure 15, however, the growth is primarily diffusion limited. The fast basal growth also contributes to the low $\sigma_{surf}$ near the tip, making quantitative analysis especially difficult. The resulting fast basal growth suggests that the SDAK mechanism is fully developed at the needle tip, and the images suggest that hollowing of the columnar tip develops as part of the growth transition.

Applying the Ivantsov parabolic solution to the growing needle tip [2019Lib], one obtains a value of $\sigma_{surf}$ that is roughly the same as that indicated in Figure 15. This assumes a far-away supersaturation of $\sigma_\infty \approx$ 1.8%, which is calculated from the experimental conditions as described in [2016Lib]. This agreement lends credence to the value of $\sigma_{surf}$ determined from the model, providing an additional check on the analysis procedure.

Pressing forward, Figure 16 shows several additional observations of the growth of columnar and dendritic structures on the ends of electric ice needles at -5 C in air. These and similar images illustrating growth at 20 different temperatures from -0.5 C to -21 C, at these same supersaturation levels, can be found in [2019Lib]. Once again, I restrict the analysis to the outer extremities of these crystals, using simply measurements of $v_{basal}$ and $v_{prism}$ growth rates.

Looking at the $\sigma_\infty = 8\%$ panel in Figure 16, the prism surfaces are large and faceted, so I use the measured $v_{prism}$ to extract $\sigma_{surf}$ at the columnar end (assuming the model in Figure 2), and then use $v_{basal}$ to extract $\alpha_{basal}$ at that $\sigma_{surf}$. The result is the second set of data points in Figure 15, at $\sigma_{surf} = 0.16\%$. As with the first set of points, this analysis suggests that the SDAK effect is fully developed on the basal surfaces, yielding a high value of $\alpha_{basal}$ on the SDAK curve. Again this is a somewhat model-dependent analysis, but the picture it paints is consistent with the model and with other observations.

Jumping ahead to the $\sigma_\infty = 128\%$ panel in Figure 16, here we find that $v_{prism}$ is only slightly lower than $v_{basal}$ at the dendrite tips, signaling a departure from the columnar behaviors at lower $\sigma_\infty$. There is little theoretical or experimental guidance from which to interpret this observation, so I speculate that $\alpha_{basal}$ remains high in this regime, meaning that $\alpha_{prism}$ must rise up to meet it as $\sigma_{surf}$ increases in Figure 15. As described at the beginning of this paper, I explain this behavior by postulating a transition to kinetic roughening on the prism surface as the growth rate increases.

To model the high-$\sigma_\infty$ data in Figure 16, therefore, I assume that $\alpha_{basal}$ roughly follows the SDAK curve in the theory (meaning $\alpha_{basal} \approx 1$ in this regime), and I use this to determine $\sigma_{surf}$ from $v_{basal}$, and then use $v_{prism}$ to extract $\alpha_{prism}$, yielding the last four data points in Figure 15. Once again, these inferences are meant to develop an overall self-consistent picture of the underlying attachment kinetics, and thus flesh out the desired kinetics model. As can be seen in Figure 15, this analysis yields a physically reasonable picture of the attachment kinetics at high growth rates, in a regime that is otherwise difficult to explore. The model is clearly highly speculative in this regime, but it can be considered as perhaps a first step toward understanding the full phenomenon of snow crystal growth dynamics. At the very least, the model suggests a framework for additional investigations and experimentation under these extreme growth conditions.

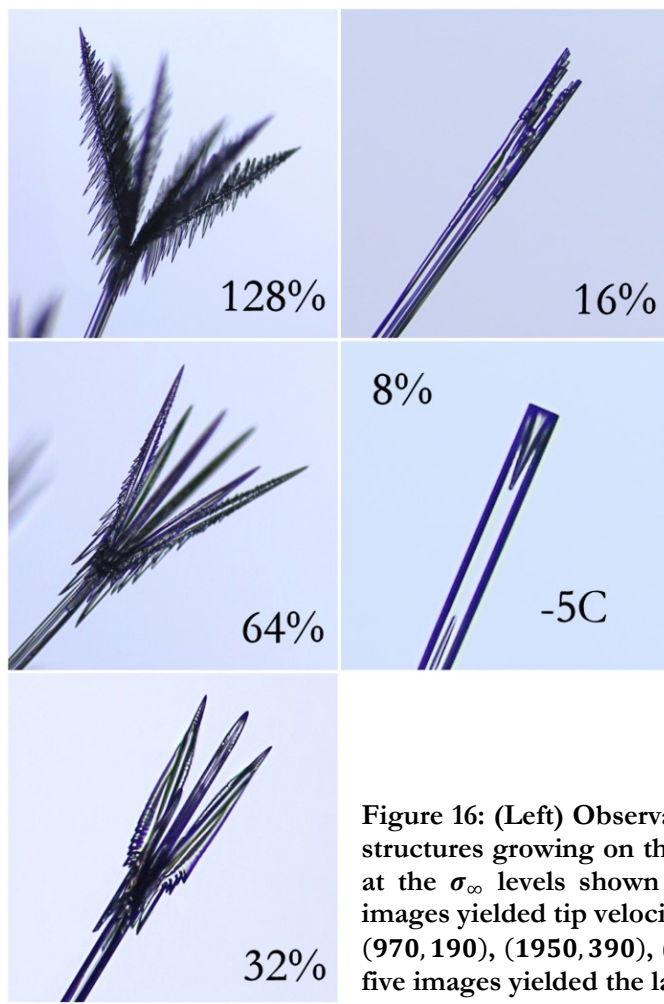

**Figure 16: (Left) Observations of the growth of columnar and dendritic structures growing on the ends of electric needle crystals at -5 C in air, at the $\sigma_\infty$ levels shown [2019Lib]. Measurements of these and other images yielded tip velocity measurements of $(v_{basal}, v_{prism}) \approx (330, 50)$, $(970, 190)$, $(1950, 390)$, $(3000, 1200)$, and $(6000, 4400)$ nm/sec. These five images yielded the last five sets of data points in Figure 15.**

### Needles and Plates in Air

For our final data set, Charles Knight recently described a series of snow-crystal growth observations at -5 C, in air at a supersaturation of roughly $\sigma_\infty \approx 4\%$ [2012Kni], finding that both plate-like crystals and slender needle-like crystals emerged in different trials. In some trials, both morphologies were seen growing simultaneously. This seemingly peculiar behavior can again be explained using the model in Figure 1, even though a quantitative analysis is not possible because the growth was so strongly diffusion limited. In essence, the apparent dichotomy of morphological behaviors can be interpreted as a manifestation of the different basal branches, as illustrated in Figure 17.

In this picture, the plate-like crystals in [2012Kni] arose when the initial growth conditions yielded relatively broad basal facets, which soon became subject to conditions of rather low $\sigma_{surf}$ as the cluster of crystals grew large. Diffusion effects quickly produced $\sigma_{surf} \ll \sigma_\infty$ around these large clusters, and in such conditions the broad basal plates retained their faceted morphology indefinitely. The low-$\sigma_{surf}$ environment no longer allowed the SDAK effect to develop away from the large-basal-surface track. Extracting crude velocity estimates from [2012Kni] places plate-like crystals at low $\sigma_{surf}$ as shown in Figure 17.

Alternatively, I propose that the slender needle crystals in [2012Kni] appeared when the SDAK effect was triggered early in the formation of the basal surfaces. Then the edge-sharpening instability [2019Lib1] caused the development of highly sharpened basal surfaces, yielding $\alpha_{basal} \approx 1$ at the needle tips. With the reported needle tip velocities of about 2 $\mu$m/sec, this implies $\sigma_{surf} \approx 0.4\%$ at the needle tip, which is again illustrated in Figure 17. Meanwhile, the bases of the needles experienced a somewhat lower $\sigma_{surf}$, thus yielding an exceedingly high $\rho_{aspect}$.

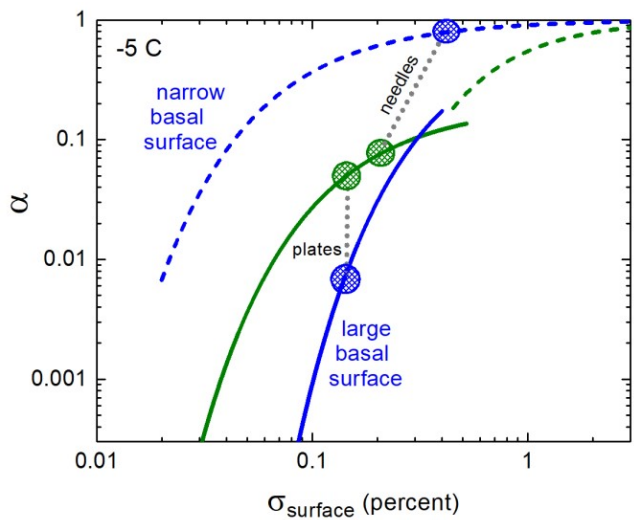

**Figure 17: An explanation for the simultaneous observation of plate-like and needle crystals in air at -5 C, as reported by Knight [2012Kni]. Diffusion effects around these large crystals give $\sigma_{surf} \ll \sigma_\infty \approx 4\%$, also producing the largest $\sigma_{surf}$ at the needle tips. Needles form if the SDAK effect is invoked via the edge-sharpening instability [2019Lib1], thus yielding $\alpha_{basal} \approx 1$ and a slender columnar morphology. Otherwise large basal surfaces appear with slow growth rates, yielding thin plates. The two markedly different morphological types thus reflect the two predominant basal tracks in our model of the attachment kinetics at -5 C.**

Again we note that $\sigma_{surf} \approx 0.4\%$ at the needle tip is not an unreasonable expectation for needle-like growth with $\sigma_\infty \approx 4\%$, as indicated by Ivantsov solution to the diffusion equation [2019Lib]. Only a rough quantitative analysis is possible here, but it supports that the observed growth behaviors are consistent with the model in Figure 1.

For both the needle and plate-like crystals, note that the overall morphology is determined mainly by the anisotropy in the attachment kinetics, and only to a lesser degree by large-scale diffusion effects. This is a general rule in snow-crystal formation that applies to essentially all growth morphologies [2019Lib]. In the case illustrated in Figure 17, $\sigma_{surf}$ is fairly low on all surfaces for both morphologies, while $\alpha_{basal}$ depends strongly on whether the crystal ends up on the SDAK track or the large-basal-surface track.

The electric-needle morphologies in the previous section are somewhat different compared to Knight's crystals, but this likely reflects the different initial conditions and crystal sizes. When the growth is strongly diffusion limited, as is the case in both these experiments, the morphology at any given time reflects the full growth history of the crystal. The electric-needle crystals are likely better suited for follow-up studies of this nature, as the initial conditions are quite well determined and reproducible [2019Lib]. However, detailed 3D numerical simulations will likely be required to fully understand all the subtle details in the formation of such complex morphological structures.

## 4. Conclusions

My overarching goal in this series of papers is to develop a comprehensive physical model of the attachment kinetics that describes the growth of ice crystals from the vapor phase. Such a model is essential for understanding the growth dynamics of snow crystals, which has been an outstanding problem for over 70 years [2019Lib]. The -5 C model depicted in Figure 1 represents a single-temperature snapshot of the general temperature-dependent model described in [2019Lib1]. A key new feature in this model is a molecular description of structure-dependent attachment kinetics (SDAK), a phenomenon that appears to play an important role in the development of complex snow-crystal morphologies over a broad range of growth conditions.

Focusing on ice growth from water vapor at -5 C, this paper examines much of the available growth data (that are sufficiently reliable) for comparison with model predictions. Although the data are limited in many respects, and the data analysis includes numerous caveats and uncertainties, overall I believe that the model does a remarkably good job of explaining the diverse collection of crystal morphologies that appear under different growth conditions at -5 C. As described above, the model reproduces growth measurements to a reasonable degree, thus providing a quantitative, albeit somewhat empirical, picture of the different physical processes involved in snow crystal formation at this temperature. Moreover, the model does a good job explaining how plate-like and columnar crystals can both commonly appear at -5 C under ostensibly identical growth conditions, as the morphological development can be strongly influenced by initial growth conditions.

Having studied this problem extensively for many years, I believe that the model in [2019Lib1] presents something of a turning point in our overall understanding of snow crystal growth. Before developing this model, I found that the various observations and growth measurements were scattered all over the map, and it was difficult to see any sense of order in the chaos. There was simply no self-consistent, physically plausible model that seemed to explain all the data. Now I feel that the pieces are beginning to fit together, as results from quite disparate experiments can all be explained to a reasonable degree. While not yet a neat and tidy package, I believe that this model represents real progress.

One clear lesson from this investigation is that it is exceedingly difficult to analyze growth data in normal air at -5 C, as diffusion effects tend to dominate the overall growth behavior and the growth velocities. The reason for this is that $\alpha_{basal}$ and $\alpha_{prism}$ are both quite high over the typically accessed range in $\sigma_{surf}$. This means $\alpha_{diff} \ll \alpha$ in many measurements, and the growth rates are therefore nearly independent of $\alpha$. This makes it impractical in many circumstances to extract useful information about $\alpha$ from growth measurements in air. The of $\alpha_{basal}/\alpha_{prism}$ ratio can be approximated from measurements in air under some conditions, but this ratio is of limited use in developing a quantitative model of the attachment kinetics.

Another important conclusion is that the SDAK effect is quite pervasive in snow crystal growth, often dominating important aspects morphological development over a broad range of growth conditions. With a physically plausible molecular model of this phenomenon as a function of temperature [2019Lib1], the overall appearance of the morphology diagram begins to make some sense, along with many quantitative growth experiments. Moreover, the SDAK phenomenon can explain many exceedingly puzzling growth behaviors, especially the appearance of different morphologies under ostensibly identical growth conditions. It appears that the somewhat capricious nature of the SDAK effect has led to much confusion in interpreting ice growth observations in the past, and hopefully this will decrease going forward.

An unfortunate feature of the SDAK phenomenon is that it becomes difficult to predict under precisely what conditions it will manifest itself. At -5 C, the prism growth is fairly well behaved, but $\alpha_{basal}$ is quite sensitive to the SDAK effect. The data presented above show that the initial growth conditions are an important factor in the development of sharp basal features, especially during periods of relatively fast growth. In the most extreme cases documented above, either thin plates or slender columns may develop under quite similar conditions, depending on the initial growth history.

One aspect of this problem is that the SDAK effect produces a bone fide edge-sharpening growth instability (ESI) [2019Lib1], and whether this instability occurs or not can depend on the entire growth history of a specific crystal. In this respect, the ESI and the branching instability share an important feature – small inhomogeneities in the growth environment may yield dramatic changes is the final crystal morphologies.

An obvious next step in this investigation is to examine ice growth data at different temperatures, building on the investigation at -5 C outlined in this paper. The model in [2019Lib1] covers a considerable range of temperatures and supersaturations, and there is hope that it might reasonably explain much of available data, and suggest many additional experimental investigations that can further probe key aspects of the model.

Another step forward will be to construct a full 3D computational model of snow crystal growth than incorporates this new model of the attachment kinetics. Our understanding of particle diffusion and surface-energy effects (mainly the Gibbs-Thomson effect) is already quite mature, so the addition of an accurate attachment-kinetics model should provide all the necessary ingredients to build a full snow-crystal simulator [2019Lib]. Progress toward this goal has been quite rapid in recent years, especially in the development of cellular-automata that already reproduce realistic-looking faceted+branched structures. It does not seem unrealistic that we may soon see an essentially complete solution of the snowflake problem, in the form of a comprehensive computational model that reproduces the full menagerie of snow-crystal structures over a broad range of input environmental conditions.